\documentclass{aa}
\usepackage{epsfig,graphicx,times}
\newcommand{\kms}{{{km~s$^{-1}$}}}
\newcommand{\teff}{{$T_\mathrm{eff}$}}
\newcommand{\logg}{{log~{\em g}}}

\newcommand{\vm}{{\em v}$_{\rm m}$\ }
\newcommand{\vi}{{\em v}$_{\rm i}$\ }

\newcommand{\vsini}{{\em v}\,sin\,{\em i}}
\newcommand{\msun}{M$_{\odot}$}
\newcommand{\lsun}{L$_{\odot}$}
\newcommand{\vrot}{$v_\mathrm{rot}$}

\begin{document}
\title{B-type Supergiants in the SMC: Rotational velocities and implications
for evolutionary models}

%\subtitle{}

\author{
P.L. Dufton\inst{1}
\and
R.S.I. Ryans\inst{1}
\and 
S. Sim\'on-D\'{\i}az\inst{2,3}
\and
C. Trundle\inst{1,3}
\and 
D.J. Lennon\inst{2,3}
} 

\institute{Department of Pure \& Applied Physics, The Queen's University 
           of Belfast, BT7 1NN, Northern Ireland, UK       
     \and
       Isaac Newton Group of Telescopes, Apartado de Correos 368, E-38700, 
       Santa Cruz de La Palma, Canary Islands, Spain
     \and
       Instituto de Astrof\'\i sica de Canarias, E-38200,
           La Laguna, Tenerife, Spain
}
\offprints{P.L. Dufton,
\email{p.dufton@qub.ac.uk}}

\date{Received  / Accepted}

\abstract{  
High-resolution spectra for 24 SMC and Galactic B-type supergiants
have been analysed to estimate the contributions of both 
macroturbulence and rotation to the broadening of their metal lines.
Two different methodologies are considered, viz. goodness-of-fit
comparisons between observed and theoretical line profiles and
identifying zeros in the Fourier transforms of the observed profiles. 
The advantages and limitations of the two methods are briefly discussed
with the latter techniques being adopted for estimating projected
rotational velocities (\vsini) but the former being used to estimate 
macroturbulent velocities.  The projected rotational velocity
estimates range from approximately 20 to 60 \kms, apart from one SMC
supergiant, Sk\,191, with a \vsini\,$\simeq$ 90\,\kms.
Apart from Sk\,191, the distribution of projected rotational 
velocities as a function of spectral type 
are similar in both our Galactic and SMC samples with larger
values being found at earlier spectral types. There is marginal evidence 
for the projected rotational velocities in the SMC being higher than 
those in the Galactic targets but any differences are only of the 
order of 5-10 \kms, whilst evolutionary models predict differences 
in this effective temperature
range of typically 20 to 70 \kms.  The combined sample is consistent
with a linear variation of projected rotational velocity with effective
temperature, which would imply rotational velocities for supergiants 
of 70 \kms\ at an effective temperature of 28\,000 K (approximately B0 
spectral type) decreasing to 32 \kms\ at 12\,000 K (B8 spectral type).
For all targets, the macroturbulent broadening would appear to be
consistent with a Gaussian distribution (although other distributions
cannot be discounted) with an $\frac{1}{e}$\ half-width 
varying from approximately 20 \kms at B8 to 60 \kms at B0 spectral types.

\vskip 1.2truecm

}

\titlerunning{B-type Supergiants in the SMC}

\maketitle

%---------------------------------------------------------- SEC: INTRO ----
\section{Introduction}

Evolutionary models, such as those by Heger \& Langer (\cite{Heg00}),
Meynet \& Maeder (\cite{Mae00}) and Maeder \& Meynet (\cite{Mae01}), 
provide predictions of the 
evolution of both surface chemical compositions and rotational velocities
of massive hot stars as a function of time. One common theme of these
models is that an initially high rotational velocity (of the order of 
300\,\kms) is required to produce significant mixing of nucleosynthetically 
processed material to the surface. Although rotational velocities decrease
as their progenitor OB-type stars evolve away from the main sequence, B-type
supergiants showing processed material at their surface are predicted
to retain significant rotational velocities. However measurement of 
projected rotational velocities for a range of SMC early-type stars 
(O-type supergiants - Hillier 
et al. \cite{Hil03}; O-type dwarfs - Bouret et al. \cite{Bou03}; B-type giants 
and supergiants - Lennon et al. \cite{Len03}, Trundle et al. \cite{Tru04})
appear to be systematically too low when compared with the predictions (see
Fig. 7 of Trundle et al.). Additionally in most previous analyses, it has 
been assumed that the line broadening was predominantly due to rotation and if
this were not the case the discrepancy between observation and theory 
would become larger. 

Howarth et al. (\cite{How97}) investigated the line widths in IUE spectra 
for a large sample of early-type evolved stars and noted that there were
no narrow-lined objects. They interpreted this as evidence that there was
another mechanism (hereafter designated as `macroturbulence'), which was
contributing to the line broadening. More recently Ryans et al.\,
(\cite{Rya02}) have analysed high signal-to-noise (S/N) ratio 
(of the order of 500) spectra of Galactic 
supergiants in order to try and estimate the relative contributions of
rotation and macroturbulence. They found that although the line widths
implied projected rotational velocities in the range of 50 to 100\,\kms,
their best estimates were typically in the range of 10 to 30\,\kms, with 
all upper limits being less than 60\,\kms. These results were lower than, for
example,  the calculations of Meynet \& Maeder  (\cite{Mey00}) for solar 
metallicity objects that predicted rotational velocities in the range of 
50 to 200\,\kms. Even allowing for projection effects, the observed 
values appeared to be systematically too low.

Recently there have been two non-LTE analyses of high quality spectroscopy
of SMC B-type supergiants. Trundle et al. (\cite{Tru04}) analysed the spectra
of 7 supergiants (and a giant) using the unified code {\sc fastwind}
(Santolaya-Rey et al. \cite{San97}) to determine their photospheric and wind
parameters. Dufton et al. (\cite{Duf04}) used the static
code {\sc tlusty} (Hubeny \cite{Hub88}; Hubeny \& Lanz \cite{Hub95}) to analyse
9 supergiants. Two of these targets were in common with those of Trundle et al. 
and a comparison of the two analyses showed excellent agreement. Here we
analyse the spectroscopic data from these two SMC investigations using the
techniques discussed in Ryans et al.\, (\cite{Rya02}) and Sim\'on-D\'{\i}az
\& Herrero\,
(\cite{Sim05b}) to estimate projected rotational and macroturbulent velocities. 
These techniques utilise different methodologies and hence allow us to 
investigate their strengths and limitations. We conclude that the Fourier
transform technique of  Sim\'on-D\'{\i}az et al.\, provides the better estimates
for the projected rotational velocity and have hence used it to reanalyse 
the Galactic stellar data presented by Ryans et al.\, (\cite{Rya02}).

Our principle aim is to obtain reliable estimates of the stellar projected 
rotational velocities while allowing for the contribution of other 
broadening mechanisms. In particular, we will investigate whether 
the low estimates found by Ryans et al.\, (\cite{Rya02}) are confirmed and
replicated in our SMC samples. One advantage of studying
targets in the low metallicity environment of the SMC is that the presence
and the amount of contamination of the stellar surface by nucleosynthetically
processed material is far easier to characterise. This in turn makes it 
easier to search for any correlation between rotation and the amount of 
such contamination.

\section{Observational data}

Our observational material consists of high resolution spectroscopy
for 13 SMC supergiants obtained at the Anglo-Australian Telescope
(AAT) and the ESO Very Large Telescope (VLT) and for 11 Galactic
supergiants obtained at the McDonald Observatory and the 
William Herschel Telescope (WHT).

Trundle et al. (\cite{Tru04}) obtained spectroscopy of 7 SMC targets
at the VLT with UVES instrument in November 2001. Of particular relevance 
here are the observations obtained in the blue arm of the spectrograph with 
a wavelength coverage of 3900 to 5000\AA. Trundle et al.\ rebinned their 
data to a pixel size of 0.2\AA\, (yielding signal-to-noise 
ratios ranging from 120 to 190). Here we have used the unbinned 
spectra which have a spectral resolution, R$\sim$40\,000. 
Two targets, AzV\,104 (B0.5Ia) and AzV\,216 (B1 III) have not been
included in the current investigation --- the former as the quality
of the observational data was insufficient to accurately estimate
the projected rotational velocity and the latter as it is not a 
supergiant. The VLT data are summarized in Table \ref{vsini_values} 
and further details of the observations and the reduction techniques 
can be found in Trundle et al. (\cite{Tru04}).

Spectroscopy for 9 SMC B-type supergiants (two in common with the
VLT observations) was obtained at the Anglo-Australian Telescope in 
September 1996, with  the UCLES instrument. There was complete wavelength 
coverage form 3900 to 4900~\AA\/ with a pixel size of $\sim$0.08~\AA,
a spectral resolution, R$\sim$20\,000 and
with signal-to-noise ratios ranging from 90 to 130. Note that although the
signal-to-noise ratios of the AAT data is lower than that of the binned VLT
data, the differences in pixel size lead to the data being of similar
quality (see, for example, Fig. 3 of Dufton et al. \cite{Duf04}). 
The stars observed are again summarized in  Table \ref{vsini_values} 
and further details of the reduction techniques can be found in Dufton 
et al. (\cite{Duf04}).

Ryans et al.\, (\cite{Rya02}) has presented spectroscopy for the 
Galactic supergiants with signal-to-noise ratios ranging from 250 to 700
and spectral resolutions of R$\sim$160\,000 (McDonald) and  
R$\sim$40\,000 (WHT). HD\,36371 (B4 Iab) has been excluded from 
this investigation due to the difficulty of estimating the projected 
rotational velocity with the other targets being listed in Table 
\ref{vsini_values}. Full details of the observations and their 
reduction can be found in Ryans et al.

\section{Analysis and Results}
\subsection{Analysis methodologies}
\label{anal_meth}
We have considered two methodologies to investigate the broadening
of the absorption lines in our supergiant sample and in particular
to distinguish between rotational broadening and that due to other
mechanisms that we characterise as macroturbulence. These utilise the
goodness-of-fit between observed and theoretical profiles and the
shape of the Fourier transforms of observed absorption lines and
are discussed in detail by Ryans et al.\ (\cite{Rya02}) and
Sim\'on-D\'{\i}az \& Herrero\ (\cite{Sim05b}) respectively. Below we 
summarize the two methodologies.

The methodology of Ryans et al.\ is, in principle, straightforward. 
Equivalent widths of relatively strong, unblended lines in each 
star are measured, and then theoretical profiles of the same strength 
are generated using a non-LTE code (e.g. {\sc tlusty} -- Hubeny 
\cite{Hub88}) with appropriate atmospheric parameters.
It is assumed that the broadening of the intrinsic absorption line
profile is due to rotation (\vsini) and another mechanism that is 
characterised as macroturbulence and is initially assumed to have a 
Gaussian distribution of velocities (with a half-width at which 
the profile has fallen to $\frac{1}{e}$ of its maximum value
given by {\em v}$_{\rm m}$). Theoretical profiles 
are then convolved with these functions and compared to the
observational data.  The sum of the squares of the differences are
computed, giving a measure of the quality of the fit at each point 
in ({\em v}$_{\rm m}$, \vsini) space. The main difficulty with this 
goodness-of-fit (GOF) approach is that profiles generated with different 
({\em v}$_{\rm m}$, \vsini) combinations are quite similar 
especially when one broadening mechanism dominates and 
we return to this point in Sect \ref{comp_meth}.

By contrast the Fourier transform (FT) methodology discussed by 
Sim\'on-D\'{\i}az \& Herrero\, (\cite{Sim05b}) allows the projected rotational 
velocity to be derived independently of any other broadening 
mechanism which may affect the line profile. A line profile 
can be considered to consist of the convolution of natural, instrumental, 
rotational and macroturbulent broadening profiles. As first discussed
by Carroll (\cite{Car33a, Car33b}) the Fourier transform of the
rotational profile has zeros whose position in frequency space
depend on the projected rotational velocity. The FT method for 
the determination of \vsini\ (c.f. Gray \cite{Gra73}) is then
based on the fact that in Fourier space, convolutions transform into
products. Hence it is possible to determine the projected rotational velocity 
of a star once the position of the first of those zeroes is identified in the
Fourier transform of the line profile. Although this methodology has 
not been widely used in the study of OB-type stars (e.g. Ebbets \cite{Ebb79}), 
the recent study by Sim\'on-D\'{\i}az \& Herrero\ (\cite{Sim05b})
illustrates the strength of this method for determining projected rotational
velocities in early type stars (see also Sim\'on-D\'{\i}az et al. 
\cite{Sim05a}).

\subsection{Comparison of methodologies}
\label{comp_meth}
We have used the high quality spectroscopy reported in Ryans et al.
(\cite{Rya02}) to compare the results obtained by the two methodologies.
We note that the FT approach does not directly provide estimates
of the macroturbulence parameter ({\em v}$_{\rm m}$) and hence the comparison 
is restricted to the estimates of the projected rotational velocities. In Table
\ref{comp_vsini}, we summarize the estimates from the two methodologies with the
goodness-of-fit results being taken directly from Ryans et al. Note that for
some lines, it was not possible to identify the zeros in the Fourier transforms
and these cases have been excluded from the comparison. Consequently the GOF
results listed in Table \ref{comp_vsini} differ slightly from those reported by
Ryans et al. The only significant change is for HD\,204172 where the exclusion
of the \ion{Mg}{ii} line at 4481\AA\, and the \ion{Si}{iii} line at 4552\AA\,
leads to an increase in the median from the GOF estimates.

\begin{table}
\begin{center}
 \caption{Comparison of the estimates of the projected rotational velocities
 found using goodness-of-fit (GOF) and Fourier transform (FT) techniques. For
 both cases, the ranges of the estimates and their medians are listed. Also 
 tabulated are the maximum possible values of \vsini\ taken directly from
 Ryans et al.\ together with the number (n) of lines considered in the current
 comparison.}
\label{comp_vsini}
 \begin{tabular}{@{}lrccrcl}
\hline
Star		& \multicolumn{3}{c}{Goodness-of-fit}
                & \multicolumn{2}{c}{Fourier Transform} & n
\\
		& Range & Median & Max. & Range & Median  \\
\hline

HD\,2905 	& 1--22  & 20 & $\leq$60 & 51--59 & 52 & 4 \\
HD\,13854 	& 11--15 & 13 & $\leq$40 & 31--34 & 32 & 3 \\
HD\,21291 	& 5--8   & 7  & $\leq$20 & 16--18 & 17 & 2 \\
HD\,24398 	& 4--19  & 9  & $\leq$40 & 34--39 & 36 & 4 \\
HD\,34085 	& 4--15  & 10 & $\leq$30 & 20--26 & 25 & 6 \\
HD\,38771 	& 5--21  & 9  & $\leq$50 & 46--55 & 50 & 4 \\
HD\,41117 	& 3--24  & 14 & $\leq$40 & 40--42 & 41 & 2 \\
HD\,193183 	& 2--47  & 20 & $\leq$50 & 40--50 & 45 & 11\\
HD\,204172	& 3--71  & 57 & $\leq$60 & 57--70 & 62 & 8 \\
HD\,206165 	& 3--41  & 11 & $\leq$40 & 25--41 & 35 & 7 \\
HD\,208501 	& 9--11  & 11 & $\leq$30 & 25--30 & 28 & 4 \\

\hline
\\
\end{tabular}
\end{center}
\end{table}

Inspection of Table \ref{comp_vsini} implies that the estimates from the
GOF technique are systematically lower than those from the FT technique with
the mean of the differences being 22$\pm$10 \kms.  However the FT 
estimates are consistent with the maximum values deduced from the GOF 
analysis. We believe that these differences reflect the difficulty in 
identifying and quantifying the relatively small amount of rotational
broadening in the presence of significant macroturbulent broadening.

With the GOF methodology, the profiles become effectively degenerate 
for small values of \vsini\ with the contour maps in the
({\em v}$_{\rm m}$, \vsini) plane having broad flat ridges along the 
\vsini\ axis (see, for example Fig. 3 of Ryans et al. \cite{Rya02}). 
Hence estimates of the projected 
rotational velocities become sensitive to the observational uncertainties 
and the assumptions adopted. For example, a Gaussian velocity distribution
has been assumed for the macroturbulence and if the distribution differed
from this, it could lead to systematic errors in the estimation of the
projected rotational velocity. Additionally the sensitivity of the GOF 
approach to uncertainties in the observational data (such as blending 
and continuum normalisation) may be the cause of the wider range of estimates 
found in a given star compared with the FT approach.

For slowly rotating stars, the Fourier transform of the broadening function
will have its first (and subsequent zeros) at high frequencies. Additionally 
if the observed stellar absorption lines are significantly broadened
by macroturbulence most of their power will occur at low 
frequencies in the Fourier domain. Since the noise in the spectrum
transforms as a `white noise' in the Fourier space (Smith \& Gray
\cite{Smi76}), then the unambiguous identification of 
the zeros in the Fourier transform becomes difficult or impossible when
the \vsini\ is low and the effect of macroturbulence is important 
(this is analogous to the difficulty in the GOF approach of estimating
projected rotational velocities when the line profiles are dominated by
macroturbulence). Indeed even for the high quality Galactic supergiant 
spectra it was not possible to estimate the projected rotational velocity 
from the Fourier transform for some spectral features in a 
given target. Tests undertaken by Sim\'on-D\'{\i}az \& Herrero\  
(\cite{Sim05b}) for cases with small projected rotational velocities 
and significant macroturbulence indicate that the former
can be overestimated by 5 to 10 \kms. Although this would reduce the
discrepancy with the estimates deduced from the GOF methodology,
it would not eliminate it. 

However the advantage of this methodology 
is that if the zeros in the Fourier Transform can be unambiguously
identified the corresponding \vsini\ estimates should be robust and in 
particular they are less dependent (compared with the GOF approach) 
on additional assumptions or on uncertainties in the observational data. 
Indeed indirect evidence for this is the smaller range in 
estimates found from different lines in a given star using the FT
methodology. 

\subsection{Estimates of the projected rotational velocity and
macroturbulence}
\label{vsini_vt}

\begin{table}
\begin{center}
 \caption{Median values for the macroturbulent (\vm) and projected rotational
 (\vsini) velocities deduced for the SMC datasets presented in Dufton et al. 
 (\cite{Duf04}) and by Trundle et al. (\cite{Tru04}). Also listed are 
 estimates for the Galactic supergiants discussed by Ryans et al. 
 (\cite{Rya02}). The sources of the spectral types and effective
 temperatures are discussed in the text.}
\label{vsini_values}
 \begin{tabular}{@{}lllrrr}
\hline
Star	& Spectral	& T$_{eff}$ & Dataset & \vsini & \vm
\\
	& type		&  K & & \multicolumn{2}{c}{km\,s$^{-1}$}   \\
\hline
\\
{\bf SMC} 
\\
AzV\,487          & BC0Ia      & 27000   & AAT      & 50 & 54 \\  
AzV\,215          & BN0Ia      & 26750   & AAT/VLT  & 66 & 55 \\
AzV\,242          & B1Ia       & 21500   & AAT      & 54 & 43 \\
AzV\,78           & B1.5Ia$^+$ & 21250   & AAT      & 46 & 36 \\
Sk\,191           & B1.5Ia     & 21500   & VLT/AAT  & 92 & 49 \\
AzV\,210          & B1.5Ia     & 20500   & VLT      & 37 & 36 \\
AzV\,462          & B1.5Ia     & 19000   & AAT      & 43 & 26 \\
AzV\,303          & B1.5Iab    & 18000   & AAT      & 42 & 26 \\
AzV\,472          & B2Ia       & 19000   & AAT      & 33 & 27 \\
AzV\,18           & B2Ia       & 19000   & VLT      & 33 & 30 \\
AzV\,374          & B2Ib       & 18500   & AAT      & 47 & 20 \\
AzV\,362          & B3Ia       & 14000   & VLT      & 38 & 30 \\
AzV\,22           & B5Ia       & 14500   & VLT      & 33 & 23 \\
\\
{\bf Galactic} 
\\
HD\,204172	& B0.2 Ia     & 28500   & WHT      & 62 & 62 \\
HD\,38771 	& B0.5 Ia     & 27500   & McDonald & 50	& 55 \\
HD\,2905 	& BC0.7 Ia    & 24000   & McDonald & 52 & 59 \\
HD\,13854 	& B1 Iab      & 23500   & McDonald & 32 & 53 \\
HD\,24398 	& B1 Ib	      & 23000   & McDonald & 36 & 38 \\
HD\,193183 	& B1.5 Ib     & 22500   & WHT      & 45 & 40 \\
HD\,41117 	& B2 Ia	      & 19500   & McDonald & 41 & 38 \\
HD\,206165 	& B2 Ib	      & 20000   & WHT      & 35 & 34 \\
HD\,34085 	& B8 Ia	      & 13000   & McDonald & 25 & 28 \\
HD\,208501 	& B8 Ib	      & 13000   & WHT      & 28 & 27 \\
HD\,21291 	& B9 Ia	      & 11500   & McDonald & 17 & 21 \\

\hline
\\
\end{tabular}
\end{center}
\end{table}

Given the discussion of the previous section, we concluded that the 
FT methodology provided the better estimates of the projected rotational 
velocities and adopted this approach for both our Galactic and SMC samples.
These estimates are summarized in Table \ref{vsini_values} together with
spectral types and estimates of the stellar effective temperature. For the
projected rotational velocities we have quoted medians as these will be less
affected by any spurious measurements although very similar values would have 
been found if we had adopted the means of the estimates. We also
undertook an analysis of the SMC dataset using the GOF methodology
and will discuss this briefly in Sect. \ref{discussion}. Spectral types 
and effective temperatures were taken from McErlean et al.\ (\cite{McE99}) 
for the Galactic sample and Trundle et al.\ (\cite{Tru04}) and 
Dufton et al.\ (\cite{Duf04}) for the SMC targets. Note that the three 
studies utilised different non-LTE codes and different
physical assumptions about e.g. the stellar winds and hence there
may be systematic differences between the effective temperature
scales. For two SMC targets, effective temperature estimates are available 
from both Trundle et al.\ and Dufton et al.\ and they differ by 500K 
and 2\,000K (in Table \ref{vsini_values} we adopt the means of these
values), which may be indicative of the magnitude of any
systematic differences.

The FT technique does not provide a direct estimate of the magnitude 
of other broadening mechanisms (that have here been characterised as 
macroturbulence) although this will affect the overall shape 
of the Fourier transform. Hence we have used the GOF technique to estimate 
this quantity. Our approach was for each star to adopt the projected 
rotational velocity estimate listed in Table \ref{vsini_values} and then 
to re-analyse the relevant lines to determine the macroturbulence 
({\em v}$_{\rm m}$) that best fitted the profile. These estimates are 
also listed in Table \ref{vsini_values} where we have again adopted the
median of the individual estimates.

The uncertainties in our estimates of the projected rotational 
velocity and macroturbulence will depend on several factors including
the quality of the observational data and the magnitude of the quantity
being estimated --- for example, an estimate of a small macroturbulence 
in a target with significant rotational broadening will have a relatively
high degree of uncertainty. In Table \ref{comp_vsini}, we list the
range of \vsini\ estimates for each of our Galactic targets with a similar
spread being found for our SMC targets. These translate into a typical 
uncertainty in the median value of the \vsini\ estimates of 10\%. The 
spread in the estimates of the macroturbulent velocity for a given target
is similar and implies a typical uncertainty in the medians of 10\%, 
apart from the smallest values (\vm$\la 30$\kms) where the uncertainty 
may be larger.

\section{Discussion} \label{discussion}
\subsection{Macroturbulence}

\begin{figure}[ht]
\includegraphics[angle=0,width=3.5in]{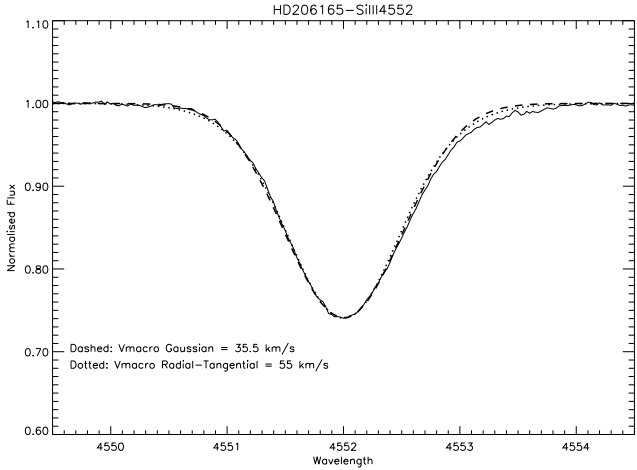}
\caption{Observed and theoretical profiles for the \ion{Si}{iii}
line at 4552\AA\ in HD\,206165. The former is represented by a histogram
whilst the theoretical profiles assume an isotropic Gaussian velocity
(dashed line) and a radial-tangential model with $A_{\rm r}=1.0$ \ 
(dotted line). Note that both theoretical approaches lead to a good fit 
with observation particularly in the core of the line, however the radially
tangential profile requires a larger macroturblent velocity.
}
\label{r_t}
\end{figure}

The macroturbulence has initially been assumed to be isotropic and to follow a
Gaussian distribution. The good agreement between the theoretical  and observed
profiles (see Fig. 1 of Ryans et al\ \cite{Rya02}) provides some evidence in
support of this assumption. However we cannot rule out other distributions and
in particular those that are qualitatively  similar to a Gaussian distribution.
As an example we consider anisotropic macroturbulence characterised by a
radial-tangential model (Gray,  \cite{Gra75, Gra92}). In this model the
velocity field is modelled by  assuming that a fraction ($A_{\rm r}$) of the
material is moving radially  (but with a Gaussian distribution of velocities)
with the remainder moving  tangentially. The methodology was developed to model
the flow of convective cells observed in the Sun. Hence it may not be
appropriate to the targets considered here and should be considered as
illustrative.  As an example we have fitted the \ion{Si}{iii} line at
4552\AA\ in  HD\,206165 adopting a radial-tangential model with 
$A_{\rm r}$=1.0 and $\Theta_{\rm r}$=55 \kms\ (i.e. assuming a complete 
radial flow). However we note that assuming that both the radial and
tangential velocities are equal ($A_{\rm r}=A_{\rm t}$, $\Theta_{\rm
r}$=$\Theta_{\rm t}$=55 \kms) would have yielded an effectively identical 
profile. In addition, we have also calculated a profile for an isotropic 
macroturbulence with $v_{\rm m}$=35.5 \kms. The results are shown in
Fig.\ \ref{r_t} and both approaches are in good agreement with the
observational data, but the radial tangential profile requires a larger
macroturbulent velocity to be invoked (by approximately 20 \kms). 
Hence we conclude that it is not possible with the current dataset to 
investigate the exact form or degree of the anisotropy of the macroturbulence. 
However what is clear is that additional broadening is present, which is 
consistent with a velocity distribution that is approximately Gaussian.

The finite spectral resolution of our observational data should not
affect our estimates of the projected rotational velocity but could 
led to overestimates for the macroturbulence. Our lowest resolution
spectroscopy (R$\sim$20\,000) would have a $\frac{1}{e}$ width, 
\vi$\sim$6 \kms assuming that the instrumental profile
was a Gaussian.  Then if the macroturbulent velocity field 
also followed a Gaussian distribution, the 
estimates listed in Table \ref{vsini_values} would represent the
instrumental profile and the actual macroturbulent velocity added in
quadrature. As an example, correcting our lowest estimate of 
\vm$\sim$ 21\kms for such an instrumental profile would lead to a 
decrease of less than 1\kms. For larger macroturbulent velocity 
estimates or better spectral resolutions, the corrections will 
be smaller and hence we have not attempted to apply them.

\begin{figure}[ht]
\includegraphics[angle=0,width=3.5in]{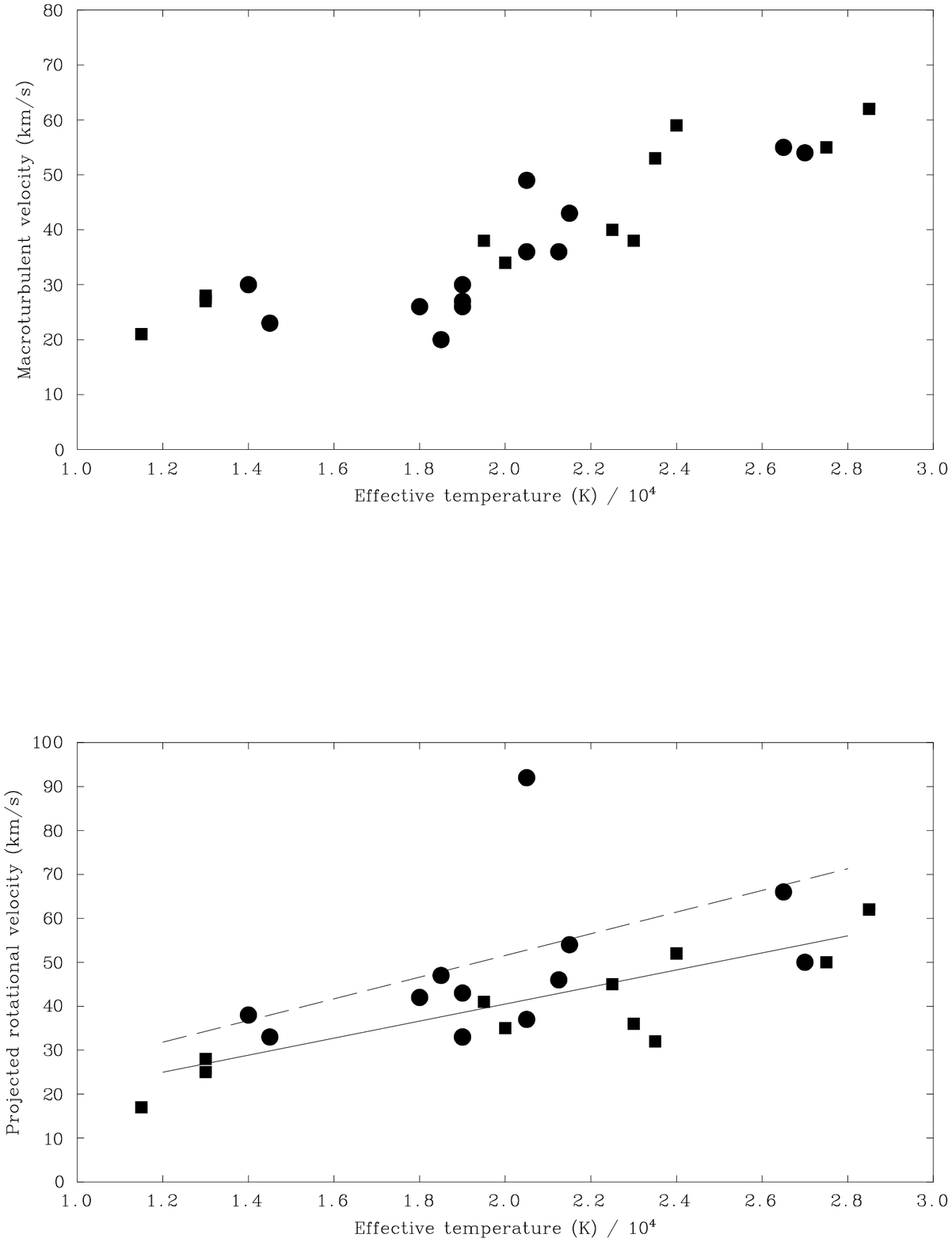}
\caption{Estimates of macroturbulent (upper figure) and 
projected rotational velocities (lower figure) plotted against effective 
temperature. The Galactic supergiants are represented by squares and the SMC 
targets by circles. For the projected rotational velocities, the linear least
squares fit (excluding Sk\,191) is shown as a solid line, whilst the dashed 
line would represent the rotational velocity assuming a random distribution of
projection angles. 
}
\label{vm0}
\end{figure}

Inspection of results summarized in Table \ref{vsini_values} implies that 
the amount of macroturbulent broadening in the Galactic and SMC supergiants 
are comparable. To illustrate this in Fig. \ref{vm0}, our best estimates 
of the macroturbulent velocities, {\em v}$_{\rm m}$, are plotted against 
the effective temperature for both sets of supergiants.
As discussed in Sect. \ref{vsini_vt}, the latter have been taken from 
different sources and although there may be systematic differences 
between these scales, we believe that this should not invalidate the 
comparison. 

In general, both samples show a similar trend of macroturbulence
with effective temperature and would imply that this is  
an intrinsic property of the star. Additionally as discussed 
previously the effect of this broadening is significant and invalidates 
the use of line widths as a measure of rotational broadening
for B-type supergiants. This is consistent with the analysis of IUE 
spectra by Howarth et al.\ (\cite{How97}), who found significant spectral
broadening in all their supergiant spectra. This they interpreted as
`confirming that an important line broadening mechanism in addition to 
rotation must be present in these objects'.

For one SMC target, Sk\,191, a relatively large projected rotational 
velocity is found from the FT methodology and this was confirmed by an
independent GOF analysis. As can be seen from Fig. \ref{vm0}, our estimate 
of its macroturbulent velocity is compatible with the other targets having 
similar effective temperatures. By contrast adopting a small rotational 
velocity would have led to an anomalously large estimate for the 
macroturbulence of approximately 80\kms. 

If the additional spectra line broadening is due to macroturbulence, the
estimated velocities, {\em v}$_{\rm m}$, are large and imply a highly
dynamic atmosphere.  Additionally B-type supergiants are also found to
require significant {\em microturbulent} velocities that need to be 
incorporated into the profiles of the line absorption coefficient, with
typically values being between 10 and 20 \kms\ (see, Crowther et al.\ 
\cite{Cro05} and references therein). If these represent small scale
turbulence, then they and the values of macroturbulence found
here imply that the photospheres of B-type supergiants are highly
turbulent on all distance scales. In turn the adoption of a stable
photosphere in both static and unified stellar atmosphere codes must be
considered problematic although excellent agreement between theory and
observation has been found using such methods (Crowther et al.\ \cite{Cro05};
Trundle et al.\ \cite{Tru04}; Dufton et al.\ \cite{Duf04}).

Assuming that our macroturbulent velocities are due to a large scale velocity
field, we can estimate its kinetic energy. As an example we consider three
relatively hot targets, HD\,38771, AzV\,215 and AzV\,487 with atmospheric
parameters, \teff$\simeq$27\,000K and \logg$\simeq$2.90. The evolutionary models
of Meynet \& Maeder (\cite{Mae00,Mae01}) then imply an evolutionary mass of
approximately 40\msun. Additionally, the corresponding {\sc tlusty} model
atmosphere implies that the mass of a column of unit area down to a Rosseland 
mean opacity of unity is 10 kgm m$^{-2}$. Adopting this as the extent of
the atmosphere sampled by the absorption lines implies a total atmospheric mass 
of $6\times10^{19}$ kgm and a typical macroturbulent velocity of 60\kms 
would then lead to a kinetic energy in the velocity field of 
approximately $10^{29}$J. It should be noted that this is only an order 
of magnitude estimate as it incorporates several assumptions and 
simplifications. However the bolometric luminosity of these supergiants is
approximately $6\times10^{5}$\lsun or $2\times10^{32}$J s$^{-1}$\ and the 
kinetic energy of the macroturbulence would be therefore negligible when
considering the energy balance in the atmosphere. McErlean et al. 
(\cite{McE98}) discuss the thermal velocities and sound speed for a
model with atmospheric parameters (\teff =27\,500 K, \logg = 3.0 and a 
microturbulence of 12 \kms) similar to those of the three targets considered 
here. For hydrogen the thermal velocity is typically 20 \kms and the sound 
speed is 15-20 \kms. Hence the energy of a macroturbulent velocity field
would be greater than the thermal energy of the plasma, whilst its velocity 
would be supersonic. Assuming that this additional broadening is indeed due 
to a large scale velocity field, its origin and nature remain unclear.

\subsection{Projected rotational velocities}

The estimates of the projected rotational velocities, summarized in Table
\ref{vsini_values}, imply that the values for both the Galactic and SMC samples
are similar and this is supported by Fig. \ref{vm0} where the estimates are 
plotted against effective temperature. If all our targets had the same 
initial mass and rotational velocity, the effective temperature would 
represent their age and evolutionary status. From a comparison with
the evolutionary models of Schaller et al. (\cite{Sch92}) for the
Galactic targets (see McErlean et al., \cite{McE99}, for further details)
and Maeder and Meynet (\cite{Mae01}) for the SMC targets (see 
Trundle et al., \cite{Tru04}), we have estimated evolutionary masses.
These fall in the range 20 to 40\msun, apart from the SMC target, 
AzV\,78, which has an estimated initial mass of approximately 50\msun.
This significant range in masses coupled with the initial rotational 
velocity of our targets being effectively unknown implies that this
figure should be considered as primarily an empirical representation.

Theoretical predictions imply that the SMC supergiants should have 
larger rotational velocities (see Fig. \ref{evol_comp} and discussion 
in Sect. \ref{evol_pred}) and there is some marginal evidence
for this in Fig. \ref{vsini_values}. To investigate this we have independently
estimated linear least squares fits for the two samples (excluding the anomalous
SMC supergiant Sk\,191 which is discussed below). The two fits are similar,
differing by 4 \kms at \teff= 12\,000K and 9 \kms at 28\,000K. Given the
small size of the samples and the possibilities of systematic differences
in the effective temperature scales of the two samples these differences
cannot be considered significant. Hence we have also combined the two samples 
and the corresponding linear fit is shown in Fig. \ref{vsini_values} as a 
solid line. The estimates of the projected rotational velocity will be 
affected by the unknown angle of inclination ($i$). Assuming that this was
randomly orientated would lead to an average value of sin\,{\em i} of 
$\frac{\pi}{4}$ and
the dashed line in the figure represents the least squares fit increased by the
reciprocal of this factor. Although the sample is small it is encouraging that
this appears to trace the upper envelope of of our estimates and would be
consistent with these stars having an angle of inclination close to 90$^o$. Hence
assuming that all B-type supergiants undergo the same evolution of rotational 
velocity (\vrot) with effective temperature our best estimate for that 
relationship is given by:

\begin{equation}
v_\mathrm{rot}= 3.4 + 2.4\times 10^{-3} T_\mathrm{eff}
\label{Eq1}
\end{equation}

It is emphasised that this relationship should be treated with
considerable caution for several reasons. Firstly our sample is relatively 
small with only 24 targets. Additionally it is unrealistic to assume that
every B-type supergiant will follow the same evolutionary history as 
this will depend on, for example, whether it is evolving on a blue 
loop having been a red supergiant 
and whether it is or was part of a binary system. Indeed the
relatively tight relationship between projected rotational velocity
and effective temperature (excluding Sk\,191) in Fig. \ref{vm0} 
is surprising. This is because a spread of estimates would be expected 
due to both the random inclination of rotational axes and to variations 
in the initial
rotational velocity. For example, targets with low \vsini\ estimates
would be expected covering the full range of effective temperatures as 
these would have evolved from main sequence O-type stars with low 
rotational velocities.

An explanation may lie in the biases in our observation samples. For example,
two supergiants were excluded from the analysis as it was not possible to
identify zeros in their Fourier Transforms, which could be due to them
having intrinsically low projected rotational velocities. Hence
our results could be biased to larger estimates, although as our initial
sample size was twenty six targets any bias would be relatively small.

For each part of our sample, the main 
criteria for target selection was as follows. For the 
Galactic supergiants, there was a large choice of possible targets and 
the final selection aimed to cover a range of spectral type for both 
Ia and Ib luminosity classes. As such there is no clear source of 
bias in our selection. The VLT spectroscopy was obtained principally
to investigate properties of the stellar winds, whilst the AAT targets
were chosen to cover a range of photospheric N/C abundances ratios
(based on the analysis of McErlean et al.\ \cite{McE99}) 
but with a bias to brighter targets for ease of observation. Hence both
SMC samples may be biased to the highest luminosity objects and indeed
only contain two objects with luminosity class Iab or Ib. Such samples
might be expected to have intrinsically small rotational velocities
due to their strong winds. However this would not explain either
the lack of low \vsini\ objects at the earliest spectra-types or the
small scatter in the estimates.

However given all the caveats, discussed above the relationship in 
equation \ref{Eq1} has some merit in indicating the magnitude of 
the rotational velocities that should be found in evolutionary models 
of blue supergiants.

\subsection{Comparison with evolutionary predictions}
\label{evol_pred}

Previously, several authors 
have estimated projected rotational velocities and compared these 
with the predictions of stellar evolutionary calculations. For the A-type
supergiants (Venn \cite{Ven99}), there appears to be reasonable agreement.
However for O-type dwarfs (Bouret et al. \cite{Bou03}), O-type supergiants 
(Hiller et al. \cite{Hil03}, Crowther et al. \cite{Cro02}), B-type giants
(Lennon et al. \cite{Len03}) and B-type supegiants (Trundle et al. 
\cite{Tru04}), the observed values of \vsini\, appear to be smaller than
those implied by stellar evolutionary calculations. However this
comparison is complicated by comparing observed {\em projected} rotational
velocities with predicted rotational velocities. Additionally
it has normally been assumed that rotation was the dominant mechanism in 
the line broadening (with for example macroturbulence not being
included) leading in effect to upper limits for the projected rotational 
velocities. The advantage of the current estimates are that we have attempted 
to distinguish between the different broadening mechanisms and thereby
have been able to estimate the actual values of projected rotational
velocities. In turn, this has allowed us to deduce a relationship between
the mean rotational velocity and effective temperature.

\begin{figure}[ht]
\includegraphics[angle=0,width=3.5in]{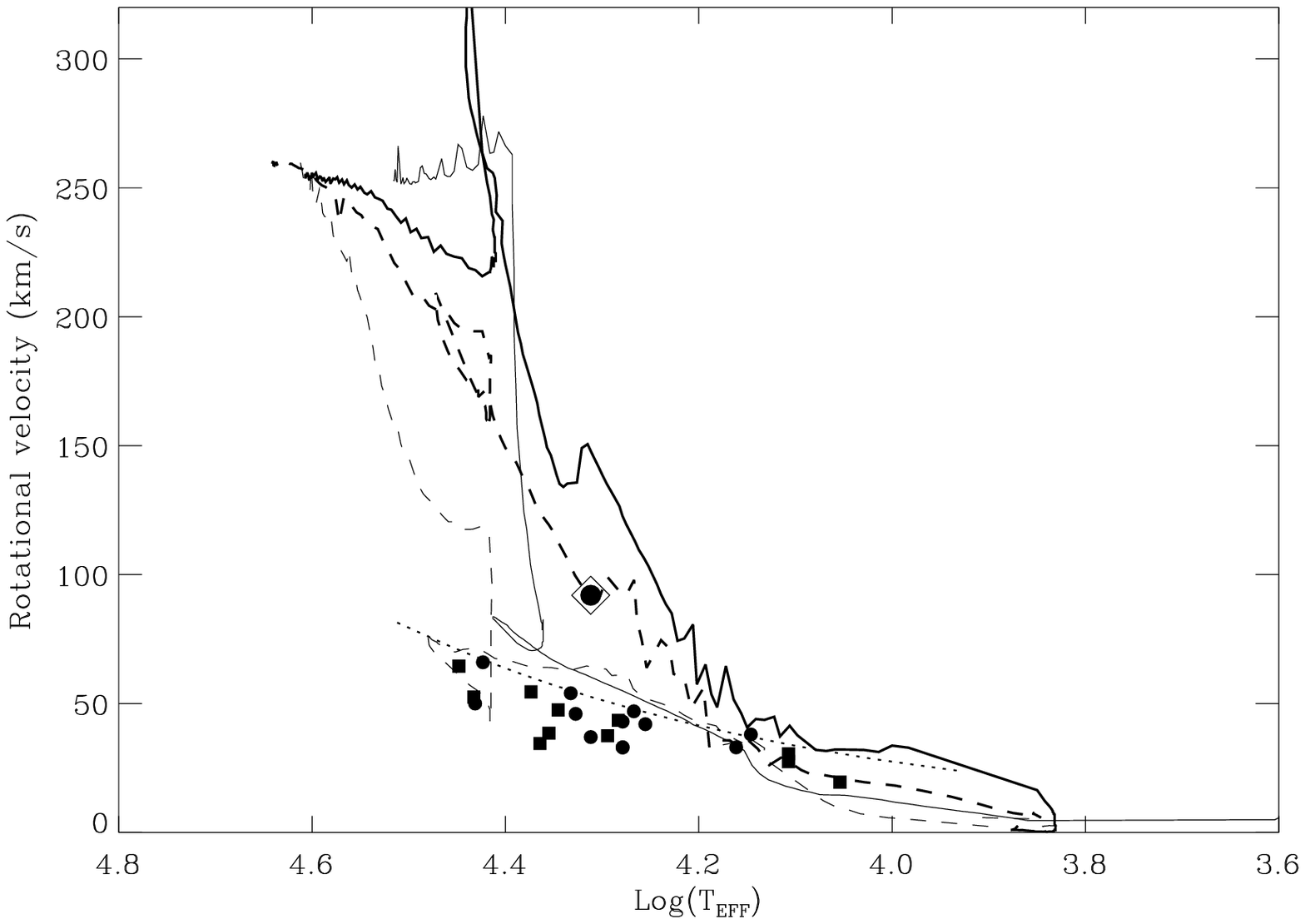}
\caption{ Predictions of the evolution of the rotational velocity of two models 
with metallicities appropriate to our Galaxy (Meynet \& Maeder \cite{Mey03};
initial masses: 20 \msun -- narrow solid line, 40 \msun -- narrow dashed line) 
and two models representative of the SMC (Meynet \& Maeder \cite{Mey05};
30 \msun -- thick solid line, 40 \msun -- thick dashed line). All models 
had an initial rotational velocity of 300\kms. Also shown are 
the estimated projected rotational velocities for our Galactic (filled squares)  
and SMC (filled circles) samples. The SMC target Sk\,191 is represented by a
filled circle enclosed in a diamond, for clarity. The dotted line represents 
our estimated mean {\it rotational velocity} for both samples, 
excluding Sk\,191 and corrected assuming a random inclination angle.
}
\label{evol_comp}
\end{figure}

In Fig. \ref{evol_comp} we compare our results with
those predicted by evolutionary models appropriate to both Galactic 
(Meynet \& Maeder \cite{Mey03}) and SMC metallicities (Meynet \& Maeder
\cite{Mey05}). All the models illustrated have initial rotational velocities
of 300 \kms and have initial masses in the range of 20 to 40 \msun. 
As the initial rotational velocities of our targets are unknown (and
indeed likely to vary from target to target), this comparison should be 
taken as illustrative. However all our SMC targets show 
nuclearsynthetically processed material in their atmospheres
and if rotation was indeed the main mechanism for this enrichment, it would
appear that their initial rotational velocities must have been relatively
large. Also shown are the measured projected rotational velocities and 
our estimate of the mean rotational velocity as a function of effective
temperature taken from equation (1). For the Galactic supergiants, our results
are in reasonable agreement with the predictions for the 40\msun\ model but may
be slightly lower than those for the 20\msun. However given the uncertainty 
in the intitial projected rotational velocities and the relatively small size 
of our sample, the agreement must be considered encouraging. The evolutionary
models appropriate to our SMC sample predict larger rotational velocities -- 
for example at log \teff = 4.4 dex, the projected rotational velocities for 
the 40\msun models are approximately 150 and 70 \kms at SMC and Galactic 
metallicities respectively. Although the projected rotational velocity of 
Sk\,191 is consistent with these predictions, those for most of the other 
SMC targets appear to be too small. This is confirmed by the discrepancy 
between the theoretical predictions and our estimate of the rotational
velocity at early B-spectral types ($ 4.45\la\log$\ \teff $\la 4.25$), although the
agreement may be better at later spectral types. We note that tests
undertaken by Sim\'on-D\'{\i}az \& Herrero\  (\cite{Sim05b}) indicate that
for relatively large macroturbulences and small projected rotational velocities,
the FT methodology may overestimate the latter. However allowing for this
effect would increase the discrepancy between theory and observation 
for the SMC dataset.

A comparison is also possible with the models of Heger \& Langer (\cite{Heg00}).
Unfortunately they do not explicitly tabulate rotational velocities as their
models evolve from the main sequence. However their calculations imply that
massive stars should evolve to higher effective temperature after their red
supergiant stage (the so called `blue loops'). Heger \& Langer predict that 
such objects should have typical rotational velocities of 50\kms\ and
that their maximum effective temperature should be approximately 15\,000K. 
For the late B-type objects, our estimated rotational velocities are smaller 
but compatible with these predictions.

\begin{figure}[ht]
\includegraphics[angle=270,width=3.5in]{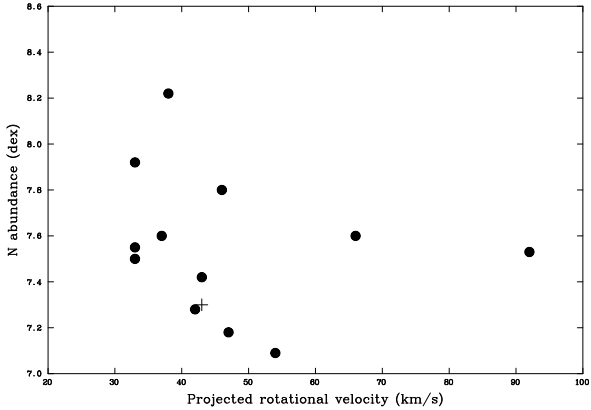}
\caption{Projected rotational velocities  plotted against nitrogen abundances 
for the SMC targets. The cross (+) is for the target, AzV\,487, for which only 
an upper limit for the nitrogen abundance could be estimated.
}
\label{n_vsini}
\end{figure}

The theoretical predictions for objects evolving from the main sequence also
imply that the amount of processed material mixed to the  surface should 
depend on the initial stellar rotational velocity. For example Maeder 
\& Meynet (\cite{Mae01}) considered 20\msun\, evolutionary  models with 
an SMC metallicity and initial rotational velocities ranging  from 0 to 
400\,\kms. At the end of the hydrogen burning phase, these models showed 
changes in their N/C abundance ratio, which were negligible for the 
non-rotating model but increased to a factor of approximately six for 
the fastest rotating model. More recently Meynet \& Maeder  (\cite{Mey05}) 
have discussed models with different initial metallicities. For models 
with significant rotational velocities, they find increases of approximately 
a factor of ten in the nitrogen abundances as the 
stars evolve into blue supergiants. 
Additionally Howarth \& Smith (\cite{How01}) have discussed
the projected rotational velocities for ON and O-type main sequence stars.
The cumulative probability function of the former implied that they
were rotating more quickly with a Kolmogorov-Smirnov test indicating
that the differences were statistically significant. Hence for these O-type
stars (that will evolve into B-type supergiants), such a correlation
between nitrogen abundances and rotational velocities appears to be 
present.

We have searched for such a correlation in our SMC sample (we do not have
reliable nitrogen abundances for our Galactic targets) by plotting 
{\em projected} rotational velocities against nitrogen abundances 
in Fig. \ref{n_vsini}. 
The latter have been taken directly from Dufton et al. (\cite{Duf04}) 
and Trundle et al. (\cite{Tru04}), although as discussed in Dufton et al. 
(\cite{Duf04}), for two stars in common there was a difference of 
approximately 0.15 dex in the nitrogen abundance estimates. Tests 
showed that allowing for this offset made no substantial difference 
and hence it has not been included in the figure. No positive 
correlation of nitrogen abundance with projected rotational velocity 
is observed. Indeed there is a suggestion that our most nitrogen
enriched targets have the lowest values of \vsini. However our sample size 
is relatively small and hence  we cannot preclude the possibilty that a 
correlation has been masked by different angles of inclination
and differences in spectral type.

Although our targets normally have relatively small rotational velocities, 
there is one exception in the SMC supergiant Sk\,191. We believe that the 
result for this star is robust as it is based on two different spectroscopic 
datasets and is supported by analyses using both the GOF and FT methodologies.
Additionally whilst the metal line profiles of the other supergiants
appear to be approximately  Gaussian, those of Sk\,191 have a profile closer 
to that characterising rotation (see, for example, Gray \cite{Gra92}). 
However the estimated atmospheric parameters and chemical composition of 
Sk\,191 are consistent with those of the other targets in our SMC sample
and hence we are unable to explain the cause of this enhanced rotation.

Hence we would appear to have a situation where effectively all our SMC  
supergiants have processed material at their surface with most 
of them having relatively small rotational
velocities. Recently, consideration has been given to the effects of
magnetic fields (Maeder \& Meynet \cite{Mae03,Mae04,Mae05}). These 
would appear to suppress mixing of nucleosynthetically 
processed material to the stellar surface. However when magnetic and
thermal instabilities are also considered,
these favour the chemical transport of elements but also lead to larger
rotational velocities during the main sequence stage of, for example,
a 15\msun model considered by Maeder \& Meynet (\cite{Mae05}).
Hence although magnetic fields may influence the presence of 
nuclearsynthetically processed material at the surface of B-type 
supergiants, current models do not appear to explain the relatively
low rotational velocity estimates found here. 

\section{Conclusions}

Our principle conclusions are as follows:
\begin{enumerate}
\item We have compared two methodologies for investigating the additional line
broadening in spectra of B-type supergiants. We find that Fourier Transforms
of the absorption line profiles yield the more reliable estimates of the 
projected rotational velocity, whilst a `goodness-of-fit' methodology can 
be used to estimate macroturbulence.
\item We have analysed AAT/UCLES and VLT/UVES spectroscopy
of 13 SMC supergiants to estimate the relative contributions of
macroturbulence and rotation to the line broadening. These have
been supplemented with a re-analysis of spectra of 11 Galactic supergiants.
\item The estimates of the macroturbulence in the SMC and Galactic supergiants
show no significant differences. Additionally both datasets show a similar 
behaviour of the macroturbulence increasing with effective temperature.
\item With the exception of Sk\,191, our estimates of the projected rotational
velocities show no significant differences between the Galactic and SMC 
targets, although the latter may be systematically larger by 5-10\kms.
As for the macroturbulence, the estimates of projected rotational
velocity increase with effective temperature.
\item Assuming a random orientation of the axis of rotation our results are
consistent with a linear behaviour between rotational velocity and effective
temperature with value of \vrot\ of 60 and 30 \kms\ at effective temperatures
of 28\,000K and 12\,000 respectively.
\item For our SMC sample, we find no positive correlation between nitrogen
abundances and projected rotational velocity as might be expected from stellar
evolutionary calculations.
\item Sk\,191 has a macroturbulence consistent with its spectral type
but an anomalously large projected rotational velocity of 92\,\kms.
\end{enumerate}

\begin{acknowledgements}
We are grateful to the staff of the Anglo-Australian Telescope and the
the European Southern Observatory for their assistance and for financial 
support from the UK Particle Physics and Astronomy Research Council.
We would like to thank Prof. I.D. Howarth for useful comments and
especially for alerting us to possible observational biases.
SS and CT are grateful for the support of the Spanish Ministerio de 
Educaci\'{o}n y Ciencia through the project AYA2004-08271-C02-01.
\end{acknowledgements}

\end{document}